\documentstyle[12pt]{article}
\begin{document}
\begin{flushright}
NTUA     50/95\\
UA-NPPS  20/95\\
Hep-Th/xxxxxxx\\
June 95\\
\end{flushright}
\vspace*{0.2cm}
\begin{center}
{\bf TACHYON FIELD QUANTIZATION}\\
{\bf   AND HAWKING RADIATION$^+$  } \\
\vspace*{0.6cm}
{\bf C. Chiou-Lahanas , G.A. Diamandis$^{*}$ , B.C. Georgalas }\\
{\bf X.N. Maintas }\\
{\it Athens University,Physics Department,}\\
{\it Nuclear and Particle Physics Section,}\\
{\it Panepistimiopolis,Kouponia,}\\
{\it GR-157 71,Athens,Greece.}\\
\vspace*{0.2cm}
{\bf and}\\
\vspace*{0.2cm}
{\bf E. Papantonopoulos$^{**}$ }\\
{\it National Technical University of Athens}\\
{\it Physics Department}\\
{\it GR-157 80 Zografou}\\
{\it Athens,Greece.}\\
\vspace*{1.0cm}
{\bf Abstact}
\end{center}
\noindent
We quantize the tachyon field in a static two dimensional dilaton
gravity black hole background,and we calculate the Hawking radiation
rate.We find that the thermal radiation flux,due to the tachyon field,
is larger than the conformal matter one.We also find that massive
scalar fields which do not couple to the dilaton,do not give
any contribution to the thermal radiation,
up to terms quadratic in the scalar curvature.

\vspace*{0.8cm}
\noindent
\thispagestyle{empty}
\vfill
\hrule
\vspace*{0.2cm}
\noindent
$^{+}$    Work partially supported by C.E.E. Science Program
SCI-CT92-0792 \\
$^{*}$  ~~ e-mail:~gdiamant@atlas.uoa.ariadne-t.gr \\
$^{**}$  ~ e-mail:~lpapa@isosun.ariadne-t.gr

\noindent
\thispagestyle{empty}
\vfill\eject
\setcounter{page}{1}

\renewcommand{\thesection}{\Roman{section}}
\section{Introduction}

Witten's identification of a black hole in string theory
\cite{Witten},has created a
considerable activity in understanding the two dimensional black hole
physics.A two dimensional dilaton gravity model coupled to free scalar
fields was proposed \cite{Callan}
as a toy model for black hole formation and
evaporation.This model enables a semiclassical treatment of Hawking
radiation and its back reaction on the geometry \cite{Russo}.

The above model can be extended to include the tachyon field,which
being a physical string mode that appears in the effective
two dimensional string theory cannot be ignored.Tachyon effects,at the
classical level,on the black hole geometry were studied
by many authors establishing the existence of two dimensional
static black hole solutions
with tachyonic hair \cite{Alwis,Kostelecky,Diamandis}.
Also a new two dimensional black hole solution with time dependent
tachyon field is discussed in \cite{Diamandis}.

It would be interesting to see the effect of the tachyon field to the
Hawking radiation of a two dimensional black hole.Kostelecky and
Perry \cite{Kostelecky},
calculating the change of the metric in the presence of the
tachyon field and taking its value on the horizon,argued that the
Hawking radiation increases in the presence of tachyonic
charge.Frolov,Massacand and Schmid \cite{Frolov}
calculated the tachyon quantum
fluctuations near the horizon,studying $<T^{2}(x)>$ in the
Hartle-Hawking vacuum,T being the tachyon field.They found that the
quantum fluctuations become very small for large black hole mass.
Nevertheless,there is no definite result in the literature
for the Hawking radiation rate coming from the contribution
of the tachyon field,as there exist for conformal matter.

We believe that a detailed study \`a l\`a Christensen and Fulling
\cite{Christensen} of the
Hawking radiation rate due to the tachyon field is interesting for two
main reasons.First of all,the tachyon field is massive,and in general
the effect of a massive scalar field to the Hawking radiation is not
well known.Secondly,the effective action emerging from the tachyon
field quantization,and in fact from the quantization of almost any
field,even if we keep terms which are quadratic in the
background fields,is highly non-local,and therefore the developing of
techniques which treat these non-localities,are interesting for their
own right.

In this work we undertake such a study,considering the one loop
effective action which emerges after the tachyon field quantization,
keeping terms which are quadratic in the background fields.
We find that the quantum effects of the tachyon field are more
significant than those of conformal matter.Also in our approximation
we find that the contribution of tachyon field to the Hawking
radiation is due to its coupling to the dilaton field.This means that
for massive scalar fields which do not couple to dilaton,the terms in
the effective action quadratic in the scalar curvature R do not
contribute to the thermal radiation of the black hole.This is a result
which is valid only in two dimensions.

In section II we set up the problem of calculating the Hawking
radiation rate,following Christensen and Fulling.Quantizing the
tachyon field,we discuss the effective action keeping terms quadratic
in the background fields.In section III we calculate
the trace of the energy momentum tensor from the effective
action.In particular we treat the non-locality using spectral analysis
of the $ \Box $ operator in the metric of a static two dimensional
black hole background.In section IV we calculate
the Hawking radiation rate due to the tachyon field.
Finally in section V we discuss our results.

\section{Effective Action}

The relation between the conformal anomaly and the Hawking radiation
is well known  \cite{Christensen}.
In particular for the
static two dimensional black hole expressed by the line element
\begin{equation}
ds^{2}=-gdt^{2}+g^{-1}dr^{2}
\label{eq:delem}
\end{equation}
where
\begin{equation}
g=1-\frac{M}{\lambda}  e^{-2 \lambda r}
\label{eq:med}
\end{equation}
the Hawking radiation rate is given by
\begin{equation}
K=\frac{1}{2} \int_{r_H}^{\infty} g^{\prime} T_{\alpha}^{\alpha} dr
\label{eq:rad}
\end{equation}
where $T_{\alpha}^{\alpha}$ is the trace of the energy
momentum tensor coming from quantum corrections and the
integral is taken from the horizon  $r_{H}=\frac{1}{2 \lambda}ln
\frac{M}{\lambda}$  to infinity. The above result is valid,provided
that the fluctuations of quantum fields in the given
background,generate a static trace anomaly.

Thus the basic ingredient for calculating the Hawking radiation rate
is the conformal anomaly.This is given by the variation of the
effective action under conformal transformations
\begin{equation}
T_{\alpha}^{\alpha}=
\frac{1}{\sqrt{\overline{g}}} \frac{\delta \Gamma[\overline{g}]}
{\delta \sigma} \mid_{\sigma =0}
\label{eq:nonam}
\end{equation}
where
\begin{equation}
\overline{g}_{\mu \nu}=e^{2 \sigma}g_{\mu \nu}.
\label{eq:conf}
\end{equation}

The classical action of the dilaton-tachyon system coupled to gravity
in two dimensions is
\begin{equation}
S=\frac{1}{2 \pi} \int d^{2}x \sqrt{-g} \{ e^{-2\Phi} [R+4(\nabla
\Phi)^{2}-(\nabla T)^{2}-V(T)+4 \lambda^{2}  ] \}
\label{eq:action}
\end{equation}
where V(T) is the tachyon potential.For the tachyon potential we take
only the quadratic part $V(T)=-m_0^2 T^2$.Redefining
$\tilde{T}=e^{-\Phi} T$ we get canonical kinetic terms for the field
$\tilde{T}$ and the action is writen
\begin{equation}
S=\frac{1}{2\pi} \int d^2 x \sqrt{-g} \{
  e^{-2 \Phi} [R+4(\nabla \Phi)^2 +4 \lambda^2]
  -(\nabla \tilde{T} )^2
  -[(\nabla \Phi)^2 -\Box \Phi -m_0^2] \tilde{T} ^2 \}
\label{eq:cact}
\end{equation}
The coupling of the tachyon to the dilaton is separated as
\[( \nabla \Phi)^2 -\Box \Phi -m_0^2=Q+m^2\]
where
     \[Q= ( \nabla \Phi)^2 -\Box \Phi -\lambda^2 \]
and
     \[m^2= \lambda^2 - m_0^2 \]
This separation is made in this way in order to recover the fact that
in the flat space with linear dilaton,the field $\tilde T $ is
massless if $ \lambda^2 = m_0^2$.Furthermore classical stability of
the background solution requires that $\lambda^2 \geq m_{0}^2$
 \cite{Diamandis,Myung}.

Quantizing the tachyon in the background of the two dimensional
static black hole,and with $\tilde{T}_{cl}=0 $,the one loop effective
action,keeping terms quadratic in the classical fields R (scalar
curvature of the metric ) and Q,is given in ref.
 \cite{Avramidi}.It consists of a local part
\begin{equation}
 \Gamma_{loc} =- \frac{1}{8 \pi}
            \int d^2 x \sqrt{g} \{ m^2 ln \frac{m^2}{\mu^2}
            +\frac{1}{2} ln \frac{m^2}{\mu^2} (Q-\frac{1}{6}R) \}
\label{eq:loc}
\end{equation}
and a non-local one
\begin{equation}
\Gamma_{nloc}=\frac{1}{8\pi} \int d^2 x \sqrt{g}
              \{Q \beta^{(1)}(\Box) Q -2Q \beta^{(3)}(\Box )R
             +R( \frac{1}{2} \beta^{(4)}(\Box)+\beta^{(5)}(\Box))R \}
\label{eq:nloc}
\end{equation}
where $\mu$ is a renormalization parameter and the operators
$\beta^{(i)}(\Box)$ are given by \cite{Avramidi}
\[ \beta^{(1)}=\frac{1}{\Box} \sqrt{\gamma}ln \frac{1+\sqrt{\gamma}}
              {1-\sqrt{\gamma}} \]
\[ \beta^{(3)}=\frac{1}{\Box} \{ \frac{\gamma -1}
              {4 \sqrt{\gamma}} ln \frac{1+\sqrt{\gamma}}
              {1-\sqrt{\gamma}} +\frac{1}{2} \} \]
\begin{equation}
 \beta^{(4)}= \frac{1}{\Box} \{
              \frac{1}{ 6 \gamma^{ \frac{3}{2} } }
              ln \frac{1+\sqrt{\gamma}}{1-\sqrt{\gamma}}
              -\frac{1}{3\gamma}
               -\frac{1}{9} \}
\label{eq:beta}
\end{equation}
\[ \beta^{(5)}=\frac{1}{48 \Box}
              \{ \{ 3-\frac{6}{ \gamma}-\frac{1}{\gamma^2} \}
              \sqrt{\gamma} ln \frac{1+\sqrt{\gamma}}{1-\sqrt{\gamma}}
              +\frac{38}{3}
              + \frac{2}{\gamma} \} \]
where
\[ \gamma=\frac{1}{1-\frac{4m^2}{\Box}}. \]
The expressions are given in Euclidean signature.

We note here that in the case Q=m=0 we get the well known result
$\frac{1}{96\pi}R \frac{1}{\Box} R$ for conformal matter which leads
to conformal anomaly $T_{\alpha}^{\alpha}=\frac{1}{24}R$.

The quadratic terms in the one loop effective action are derived in
\cite{Avramidi} for any dimension.Especially in two dimensions the
diagrammatic derivation of the above non-local terms is simple.In
particular the $Q^2$ term comes from a finite graph.This graph
can be evaluated for example in flat space and then the appropriate
covariantization yields the result in (\ref{eq:beta}).Furthermore the
QR term comes from the contribution of the tadpole graph in the
effective action and can be calculated in the same way,as the $Q^2$
one.The $R^2$ terms can also be derived diagrammatically in the
light-cone gauge.The diagrammatic approach can also used for the
derivation of the higher order one-loop terms.The diagrammatic
derivation of the form of the effective action is beyond
the scope of this paper,so it
is not given here.Details for the calculation and covariantization
along the above lines,can be found in \cite{Georgalas}.

\section{The Trace of the Energy Momentum Tensor}

The non-local character of the terms in (\ref{eq:nloc}),does not
permit a straightforward variation of the action with respect to the
metric.In order to deal with the non-local terms we use spectral
analysis of the $\Box$
operator in the given background.We work in the "unitary gauge" where
the metric of the black hole has the wellknown cigar form
\cite{Witten}
\begin{equation}
ds^2= \frac{1}{4 \lambda^2} tanh^2(\lambda y) d\theta^2
      +d y^2
\label{eq:cigar}
\end{equation}
and the angle coordinate $\theta$ is the compactified Euclidean
time.The eigenvalue equation for the $\Box$ operator is
\begin{equation}
\frac{4 \lambda^2}{tanh^2( \lambda y) }
\frac{\partial^2 \Psi}{\partial \theta^2}
+\frac{1}{tanh( \lambda y) } \frac{\partial}{\partial y}
(tanh( \lambda y) \frac{\partial \Psi}{\partial y} )
+ \tilde{\kappa}^2 \Psi = 0
\label{eq:eigv}
\end{equation}
Since the background fields are static quantities we
ignore the $\theta$ dependence and we look for spherically symmetric
solutions,satisfying the equation
\begin{equation}
g^{-\frac{1}{2}} \frac{d}{dx}
[g^{\frac{1}{2}} \frac{d \Psi_{\kappa} }{dx} ]
+\kappa^2 \Psi_{\kappa}  =0
\label{eq:geq}
\end{equation}
where we have defined $g^{\frac{1}{2}}=tanh \frac{x}{2}$  $,$
$x=2 \lambda y$ , and $\kappa^2=\frac{\tilde{\kappa}^2}
{4 \lambda^2}$.

The solution of the above equation,which is regular on the horizon
$(x=0)$ and with a plane wave asymptotic behavour $(x \rightarrow
\infty)$, is
\begin{equation}
\Psi_{\kappa} (x)= N(\kappa) _{2}F_{1}
 (i\kappa,-i\kappa,1;-sinh^2 \frac{x}{2} )
\label{eq:psi}
\end{equation}
where the normalization constant $N(\kappa)$ is given by
\begin{equation}
N(\kappa)= \frac{1}{\sqrt{2\pi}} | \frac{\Gamma(i\kappa)
            \Gamma(1+i\kappa) }
            {\Gamma(2 i \kappa ) } |
\label{eq:norm}
\end{equation}
The eigenfunctions satisfy the orthogonality relations
\[ \int_{0}^{\infty} dx \sqrt{g(x)} \Psi_{\kappa} (x) \Psi_{\mu} (x)
   =\delta(\kappa - \mu) \]
and the completeness relation in the $x$ direction
\[ \int_{0}^{\infty} d \kappa \Psi_{\kappa} (x) \Psi_{\kappa}(z)
 =\delta^{(1)}(x,z) = \frac {\delta(x-z)}{\sqrt{g(x)}} \]
Any static scalar quantity $H(x)$ can be expanded as
\begin{equation}
H(x)= \int_{0}^{\infty} d \kappa C_{H} (\kappa) \Psi_{\kappa}(x)
\label{eq:he}
\end{equation}
where
\begin{equation}
C_{H}(\kappa) =\int_{0}^{\infty}
 dx \sqrt{g(x)} H(x) \Psi_{\kappa}(x)
\label{eq:cih}
\end{equation}
In particular the scalar curvature is given by
\begin{equation}
R(x)=4 \lambda^2 \int_{0}^{\infty} d\kappa C(\kappa) \Psi_{\kappa}(x)
\label{eq:pho}
\end{equation}
with
\begin{equation}
C(\kappa)=N(\kappa) | \Gamma (1+i\kappa) |^2
\label{eq:ci}
\end{equation}
Using the previously described spectral analysis,the non-local part of
the effective action (\ref{eq:nloc}) becomes

\[   \Gamma_{nloc}= \frac{1}{8 \pi}
           \int_{0}^{\infty} d \kappa [\frac{1}{4\lambda^2}
            C_{Q}(\kappa) C_{Q}(\kappa) \beta^{(1)}(\kappa)
           -2 C_{Q}(\kappa) C (\kappa) \beta^{(3)}(\kappa) \]
\begin{equation}
            + 4 \lambda^2 C (\kappa) C (\kappa) \{
      \frac{1}{2}\beta^{(4)}(\kappa)+ \beta^{(5)}(\kappa) \} ]
\label{eq:efact}
\end{equation}
where $\beta^{(i)}(\kappa)$ are given in (\ref{eq:beta}) with the
substitution $ \Box \rightarrow  -4 \lambda^2 \kappa^2 $,
and
 \[ C_{Q}(\kappa) = \int_{0}^{\infty}
dx \sqrt{g(x)}Q(x) \Psi_{\kappa}(x) \]
Note that if $m\neq 0$ then the functions $\beta^{(i)}(\kappa)$ have a
finite limit at $\kappa =0$.

According to our previous discussion,we need to calculate the trace of
the energy momentum tensor.In order to derive it from
 the effective action written in the form
(\ref{eq:efact}),we need the variation of C and $C_{Q}$ under conformal
transformations.
Eq.(\ref{eq:cih}) implies that the variation of the expansion
coefficients can be derived from the variation of the corresponding
scalar function,which is easy to perform,and the variation of the
eigenfuctions of the $\Box$ operator.
For this it is more convenient to work with the rescaled functions
\begin{equation}
U_{\kappa}=g^{\frac{1}{4}} \Psi_{\kappa}(x)
\label{eq:res}
\end{equation}
which satisfy,according to (\ref{eq:geq}),a
Schr\"odinger-like equation
\begin{equation}
\frac{d^2}{dx^{2}}U_{\kappa}(x)+
   [\kappa^2 -V(x) ] U_{\kappa}(x) =0
\label{eq:schr}
\end{equation}
where
\[ V(x)=\frac{1-2coshx}{4sinh^2 x} \]

If we perform the conformal transformation (\ref{eq:conf})
with the conformal factor depending only on $x$, the eigenfunctions
(\ref{eq:res}) become
\[ \overline{U}_{\kappa}=\overline{g}^{ \frac{1}{4}} \overline{\Psi}
  _{\kappa}(x) \]
which will satisfy eq.(\ref{eq:schr}) with
\[ \overline{V}(x)=V(x) - \kappa^2(e^{2\sigma (x)}-1) \]
Since we are interested at the variation of U's at the point
$\sigma=0$,as we can see from (\ref{eq:nonam}),
we can solve the equation (\ref{eq:schr}) perturbatively,getting
\begin{equation}
\overline{U}_{\kappa}(x)=U_{\kappa}(x)
    -\int _{0}^{\infty}d \mu \frac
     {<U_{\mu} |2 \kappa^2 \sigma| U_{\kappa} >}
     {\kappa^2 - \mu^2 } U_{\mu}(x) + O(\sigma^2)
\label{eq:pert}
\end{equation}
where using (\ref{eq:res}) we have
\begin{equation}
 <U_{\mu} |2 \kappa^2 \sigma| U_{\kappa} >=
  \int_{0}^{\infty} dx g^{\frac{1}{2}}(x) 2 \kappa^2
  \sigma(x) \Psi_{\mu}(x) \Psi_{\kappa}(x)
\label{eq:ele}
\end{equation}
Then the eigenfunctions read
\begin{equation}
\overline{\Psi}_{\kappa}(x)=\Psi_{\kappa}(x)
    -\int _{0}^{\infty}d \mu \frac
     {<\Psi_{\mu} |2 \kappa^2 \sigma| \Psi_{\kappa} >}
     {\kappa^2 - \mu^2 } \Psi_{\mu}(x) + O(\sigma^2)
\label{eq:perps}
\end{equation}
Their variation,keeping the first order term in the
perturbation series, is given by
\begin{eqnarray}
 \frac{\delta \overline{\Psi}_{\kappa} (x) }
        {\delta \sigma (y) } |_{\sigma=0} &=&
        -2 \kappa^2 \Psi_{\kappa}(y)
        \int_{0}^{\infty} d \mu \frac
        {\Psi_{\mu}(y) \Psi_{\mu}(x) }
        {\kappa^2 - \mu^2 } \nonumber \\
   &=& -2 \kappa^2 \Psi_{\kappa}(y) G_{\kappa}^{(1)}(y,x)
\label{eq:dy}
\end{eqnarray}
where $G_{\kappa}^{(1)}(y,x) $ is the Green function satisfying
\begin{equation}
(\Box_{y} +\kappa^2) G_{\kappa}^{(1)}(y,x)=
         \delta^{(1)} ( y,x)=\frac{\delta(y-x)}{\sqrt{g(y)} }
\label{eq:grf}
\end{equation}
Using now the relation (\ref{eq:cih}) we find
\begin{equation}
\frac{\delta C(\kappa)}{\delta \sigma(x)} |_{\sigma=0} =
  +2 \kappa^2 \sqrt{g(x)} \Psi_{\kappa}(x)
 -2 \int d \mu C(\mu) \Psi_{\kappa} (x) \Psi_{\mu}(x) \\
 \frac{\kappa^2}{\kappa^2 -\mu^2} \sqrt{g(x)}
\label{eq:doolc}
\end{equation}
\begin{equation}
\frac{\delta C_{Q}(\kappa)}{\delta \sigma(x)} |_{\sigma=0} =
  -2 \lambda^2 \sqrt{g(x)} \Psi_{\kappa}(x)
 -2 \int d \mu C_{Q}(\mu) \Psi_{\kappa} (x) \Psi_{\mu}(x) \\
 \frac{\kappa^2}{\kappa^2 -\mu^2} \sqrt{g(x)}
\label{eq:delcq}
\end{equation}

Note that the described procedure can be applied
in the general case
including
the $ \theta $  dependence of the eigenfunctions
and considering conformal transformations,
where the conformal factor is
$\sigma(x,\theta) $.Using eqs.(\ref{eq:doolc}) and (\ref{eq:delcq})
we can write down the trace of the energy momentum tensor
\begin{eqnarray}
T_{\alpha}^{\alpha} =&-& \frac{4\lambda^4}{8 \pi}   \{
  \int_{0}^{\infty} C(\kappa) [\beta^{(1)}(\kappa)-4 \beta^{(3)}(\kappa)]
    \Psi_{\kappa}(x)d \kappa \nonumber \\
  &+& \int_{0}^{\infty} d \kappa C(\kappa)
  [\beta^{(1)}(\kappa)-4 \beta^{(3)}(\kappa)]
\int _{0}^{\infty} d \mu C(\mu)
\Psi_{\kappa}(x) \Psi_{\mu}(x) \frac{\kappa^2}{\kappa^2-\mu^2}
 \nonumber \\
  &+&  \int _{0}^{\infty} d \kappa C(\kappa)
  [16 \beta^{(t)}(\kappa)- 4\beta^{(3)}(\kappa)]
 \int _{0}^{\infty} d \mu C(\mu) \Psi_{\kappa}(x) \Psi_{\mu}(x)
  \frac{\kappa^2}{\kappa^2-\mu^2}  \nonumber \\
 &-&  \int_{0}^{\infty} C(\kappa)\kappa^2
 [16 \beta^{(t)}(\kappa)- 4\beta^{(3)}(\kappa)]
    \Psi_{\kappa}(x) d \kappa \}
\label {eq:trace}
\end{eqnarray}
where $\beta^{(t)}=\frac{1}{2}\beta^{(4)}+\beta^{(5)}$.
In the derivation of the (\ref{eq:trace}),we have used the fact that
in the the static black hole background,$Q=\frac{1}{4} R$ which
implies that $C_{Q}(\kappa)=\lambda^2 C(\kappa)$.This relation can be
used only after the variation,since the two functions behave
differently under conformal transformations as can be seen from eqs.
(\ref{eq:doolc}) and (\ref{eq:delcq}).

The above expression is quite complicated and especially the terms
involving double momentum integration cannot be brought to a simpler
expression in configuration space, for the specific form of the
functions $ \beta(\Box) $ in eqs.(\ref{eq:beta}).Nevertheless one can
easily check that for local operators and for the oparator
$\frac{1}{\Box}$ which appears in the one loop effective
action of the conformal matter,the above expression reproduces
the familiar results.In particular for the operator $\frac{1}{\Box}$
the double momentum integral vanishes for symmetry reasons,and thus a
trace anomaly proportional to R arises from the effective action of
conformal matter.

\section{Hawking Radiation Rate}

The trace of the energy momentum tensor retains its form if we go to
Minkowski signature,and if we use the metric of eq.(\ref{eq:med}),
the only change will be in the arguments of the hypergeometric
functions in (\ref{eq:psi})
\begin{equation}
\Psi_{\kappa} (r)= N(\kappa) _{2}F_{1}
 (i\kappa,-i\kappa,1;1-\frac{\lambda}{M}e^{2 \lambda r} )
\label{eq:psir}
\end{equation}
Now using (\ref{eq:trace}) the Hawking radiation rate in
(\ref{eq:rad}),becomes
\begin{eqnarray}
    K  = & - &  \frac{\lambda^4}{4}
  \int_{0}^{\infty} d\kappa \int _{0}^{\infty} d t C(\kappa) \{
  [\beta^{(1)}(\kappa)-4 \beta^{(3)}(\kappa)]  \nonumber \\
  & - & \kappa^2  [16\beta^{(t)}(\kappa)- 4\beta^{(3)}(\kappa)] \}
 e^{-t}N(\kappa) _{2}F_{1}(i\kappa,-i\kappa,1;1-e^t)
 \nonumber \\
 & - &  \frac{\lambda^4}{4}
  \int_{0}^{\infty} d\kappa \int_{0}^{\infty}d \mu
 \int _{0}^{\infty} d t C(\kappa) C(\mu) \{
  [\beta^{(1)}(\kappa)-4 \beta^{(3)}(\kappa)]  \nonumber \\
  & + &   [16\beta^{(t)}(\kappa)-4 \beta^{(3)}(\kappa)] \}
\frac{\kappa^2}{\kappa^2-\mu^2} N(\kappa) N(\mu) \nonumber \\
&  & e^{-t}~ _{2}F_{1}(i\kappa,-i\kappa,1;1-e^t)
_{2}F_{1}(i\mu,-i\mu,1;1-e^t)
\label {eq:tracef}
\end{eqnarray}
The expression for K consists of two terms.The first term involves a
space integration (t-integration) which can be easily performed
giving,
\[ \int _{0}^{\infty} e^{-t}~ _{2}F_{1}(i\kappa,-i\kappa,1;1-e^t)
   =|\Gamma(1+i\kappa)|^2 , \]
and a $\kappa$ integral that can be calculated
analytically or
numerically for any form of the functions $\beta(\kappa)$.
The second term is much more complicated reflecting the high
non-locality of the effective action,involving the space integration
and a double integral over the momentum.The interesting thing here
is that the integral  over the space coordinate and the
$\mu$-momentum
variable,can be evaluated analytically.To see this we
consider
\begin{eqnarray}
 &   & \int_{0}^{\infty}  d\mu \int_{0}^{\infty}dt e^{-t} C(\mu)
  \frac{1}{\kappa^2 - \mu^2} \Psi_{\mu}(t) \Psi_{\kappa}(t)
  \nonumber \\
   & = &\int_{0}^{\infty} d\mu \int_{0}^{\infty}dt e^{-t}
  \int_{0}^{\infty}dz \sqrt{-g(z)}R(z)
  \frac{1}{\kappa^2 - \mu^2} \Psi_{\mu}(t) \Psi_{\kappa}(t)
 \Psi_{\mu}(z)   \nonumber \\
 & = & \int_{0}^{\infty} dz \int_{0}^{\infty}dt e^{-t}
 \sqrt{-g(z)}R(z) G_{\kappa}^{(1)}(z,t)  \Psi_{\kappa}(t)
\label{eq:calc}
\end{eqnarray}
with $G_{\kappa}^{(1)} (z,t) $ the Green function defined in
(\ref{eq:dy}).
We can proceed further and write
\begin{equation}
\int_{0}^{\infty}dz \int_{0}^{\infty}dt e^{-t} e^{-z}
 G_{\kappa}^{(1)}(z,t)  \Psi_{\kappa}(t)
=\int_{0}^{\infty} dz e^{-z} F_{\kappa}(z)
\label{eq:fos}
\end{equation}
where
\[ F_{\kappa}(z)=\int_{0}^{\infty} e^{-t} \Psi_{\kappa}(t)
   G_{\kappa}^{(1)}(z,t) \]
a field satisfying the equation
\begin{equation}
(\Box_{\kappa} + \kappa^2) F_{\kappa}(z)=
                e^{-z} \Psi_{\kappa}(z)
\end{equation}
The above equation can be solved,using Laplace transformations,
yielding for the field $ F_{\kappa} (z)  $
\begin{equation}
F_{\kappa}(z)=
N(\kappa)(e^z-1)_{2}F_{1}(1-i\kappa,1+i\kappa,2;1-e^z)
\label{eq:fs}
\end{equation}
Inserting the form of the field $ F_{\kappa}(z) $ in (\ref{eq:fos})
we have
\begin{equation}
\int_{0}^{\infty}dz \int_{0}^{\infty}dt e^{-t} e^{-z}
 G_{\kappa}^{(1)}(z,t)  \Psi_{\kappa}(t)
= N(\kappa)|\Gamma(1+i\kappa)|^2
\label{eq:fres}
\end{equation}
So the second term in (\ref{eq:tracef}) reduces to a single
$\kappa$-integration
\begin{eqnarray}
&-&\frac{\lambda^4}{4} \int_{0}^{\infty}d\kappa C(\kappa)
N^2(\kappa)\{
[ \beta^{(1)}(\kappa)-4\beta^{(3)}(\kappa)] \nonumber \\
&+& \kappa^2 [ 16\beta^{(t)}(\kappa)-4\beta^{(3)}(\kappa) ]
 |\Gamma (1+i\kappa)|^2 \}
\label{eq:douit}
\end{eqnarray}
Combining eq.(\ref{eq:douit}) with the $\kappa$-integration from the
first term the final expression for the Hawking radiation rate
becomes
\begin{equation}
K= -\frac{\lambda^2}{16} \int_{0}^{\infty}d\kappa C(\kappa) N(\kappa)
(\kappa^2 +1)
 \{ \overline{\beta}^{(1)}(\kappa)-4 \overline{\beta}^{(3)}(\kappa) \}
 |\Gamma (1+i\kappa)|^2
\label{eq:hawra}
\end{equation}
where
\[\overline{\beta}^{(i)}(\kappa)= 4 \lambda^2 \overline{\beta}
^{(i)}(\kappa) \]

Thus we see that the treatment of the non-local action has led to a
simple expression for the Hawking radiation rate of the static two
dimensional black hole.This turns out to be independent of the mass of
the black hole as expected in two dimensions,and depends on the mass
of the tachyon field,and especially on the parameter $
\tilde{m}^2=\frac{m^2}{\lambda^2} $.For the particular forms of the
functions $ \beta(\kappa) $, the $ \kappa $-integral cannot be
evaluated analytically,but it can be calculated numerically.In figure
1 we give the plot of the ratio of the Hawking radiation rate
due to the tachyon,over the one due to conformal matter,
as a function of $\tilde{m}^2$.

\section{Conclusions and Discussion}

The numerical calculation of the Hawking radiation rate shows that the
tachyon field radiation is enhanced with respect to the conformal
matter one as it is shown in Fig.1.
This is slightly above $\frac{\lambda^2}{48}$ at
the maximum value of the mass of the tachyon field $(\tilde{m}^2
\rightarrow 1)$ but becomes significantly larger at lower tachyon
masses.As the tachyon mass tends to zero the rate tends to
infinity,because of the infrared divergence.This divergence is due to
the presence of the Q field which even in the absence of mass gives
deviation from the conformal coupling.In particular the term
responsible for the divergence is of the form
$Q\frac{1}{\Box}ln(-\frac{\Box}{m^2})Q$ as can be seen from a small
mass expansion of the effective action.

If we were dealing not with a tachyon field but with a field with
ordinary mass term,and the same coupling with the dilaton,
then $\tilde{m}^2= \frac{m_{0}^2 +\lambda^2}
{\lambda^2}$.In this case the parameter  $\tilde{m}^2$ can be large
independently of $\lambda^2$ and we can see that the Hawking radiation
rate tends to zero as the mass becomes large.This is expected since
such a massive field is actually classical.

One interesting feature that comes out from our calculation is that in
the background of the static two dimensional black hole,if the
effective action is just of the form $R \beta(\Box,m^2) R $ with
$\beta(\Box,m^2)$ any operator with regular zero momentum limit,
then the rate of the thermal Hawking radiation is zero,
except from the case of conformal matter,where $\beta=\frac{1}{\Box}$,
which does not satisfy the regularity requirement but
yields the wellknown result.This feature can be
directly tested for local actions,without involving the analysis
adopted in this work,but as we have shown it can be extented to all
non-local actions except the one for the conformal matter.In all these
cases any signal coming out from the horizon of the black hole,
due to these terms, must be of non-thermal type.From this result,which
is a peculiarity of the two dimensional black hole background,one
cannot infer that in the presence of terms of this type only,the black
hole is quantum mechanically stable,without studying the back-reaction
effects.For the tachyon field the enhancement of the thermal
radiation is exclusively due to its coupling with the dilaton.

The one loop effective action,unlike the case of conformal
matter,has also higher order terms (cubic,quartic,...)
in the background fields Q and R.
These terms have an extra  ($\frac{1}{\Box} $) non-locality,
as it is easily seen from dimensional arguments,at next order.
For example $ R \frac {1}{ \Box } R \frac{1}{ \Box } R $
is expected
to give the order of magnitude for the cubic terms.
The equation $ \Box \Phi = R $ has the
solution $ \Phi =x =2 ( \Phi_{H} - \Phi ) $ ,
where $ x $ is the dimensionless
coordinate variable we have used previously and $ \Phi $
the dilaton field,while
$ \Phi_{H} $ is its value on the horizon.
Using this prescription of the non-locality the
quadratic terms are of the order $ x e^{-x} $ and the
qubic terms are of order $ x^2 e^{-x} $.
We see that near the horizon
($ x=0 $) the quadratic terms are more significant than
the qubic ones and the same holds for their contribution
to the trace of the energy momentum tensor.
{}From eq.(\ref{eq:tracef}) we see,that the main contribution to the
Hawking radiation rate,comes from the region near the horizon,
due to the form of the conformal anomaly and the
factor $ g^{'} $ .Thus we conjecture that the higher
order terms do not alter the results
significantly.Of course this is not a proof and a more
solid answer to
this question requires direct calculation of the
contribution of the
next order terms.The method described in this work
can be used for such a calculation.

\newpage
\noindent
{\bf FIGURE CAPTIONS}

\vspace{1cm}
\parindent=-1.5cm
{\bf Fig.1:}
The plot of the ratio of the Hawking radiation rate due to the tachyon
over the one due to conformal matter,as a function of
$\tilde{m}^2$.

\end{document}